\documentclass[sort&compress,final]{aipproc}

\layoutstyle{6x9}

\usepackage{graphicx}
\usepackage{bm}

\newcommand{\beq}{\begin{equation}}
\newcommand{\eeq}{\end{equation}}
\newcommand{\bqa}{\begin{eqnarray}}
\newcommand{\eqa}{\end{eqnarray}}

\def\simge{\mathrel{
    \rlap{\raise 0.511ex \hbox{$>$}}{\lower 0.511ex \hbox{$\sim$}}}}
\def\simle{\mathrel{
    \rlap{\raise 0.511ex \hbox{$<$}}{\lower 0.511ex \hbox{$\sim$}}}}
    
\begin{document}

%%%%%%%%%%%%%%%%%%%%%%%%%%%%%%%%%%%%%%%%%%%%
%% Title and Abstract
%%%%%%%%%%%%%%%%%%%%%%%%%%%%%%%%%%%%%%%%%%%%

\title{Thermal Bottomonium Suppression}

\classification{11.15Bt, 04.25.Nx, 11.10Wx, 12.38Mh}
\keywords{Quarkonium Suppression, Relativistic Heavy Ion Collision, Quark-Gluon Plasma}

\author{Michael Strickland}{
  address={Physics Department, Gettysburg College, Gettysburg, PA 17325 United States}
}

\begin{abstract}
I discuss recent calculations of the thermal suppression of bottomonium states in 
relativistic heavy ion collisions.  I present results for the inclusive $\Upsilon(1s)$
and $\Upsilon(2s)$ suppression as a function of centrality. 
I compare with recent CMS preliminary 
data available at central rapidities and make predictions 
at forward rapidities which are within the acceptance of the ALICE dimuon spectrometer.
\end{abstract}

\maketitle

%%%%%%%%%%%%%%%%%%%%%%%%%%%%%%%%%%%%%%%%%%%%
%% Text
%%%%%%%%%%%%%%%%%%%%%%%%%%%%%%%%%%%%%%%%%%%%

%---------------------------------------------------------------------------------------------------------------
\section{Introduction}
%---------------------------------------------------------------------------------------------------------------

The purpose of ongoing and upcoming heavy ion collision experiments at the Relativistic Heavy 
Ion Collider (RHIC) and the Large Hadron Collider (LHC) is to study the behavior of nuclear 
matter at high energy density, $\epsilon \gg 1\;{\rm GeV/fm}^3$.  At such high energy 
densities one expects to create a deconfined quark gluon plasma (QGP).  One of the key observables 
measured in the experiments is the relative yield of heavy quarkonium production relative to
that in proton-proton collisions scaled by the number of nucleon collisions, $R_{AA}$.  Early 
studies predicted that at high temperatures Debye screening
would lead to the dissolution of hadronic bound states \cite{Shuryak:1980tp}.  Therefore, one
might be able to use the relative suppression of heavy quark bound states as a ``smoking
gun'' for the creation of the quark gluon plasma.  Over the last 25 years most of the interest 
has been focused on the suppression of heavy quark-antiquark bound states following early
predictions of thermal $J/\Psi$ suppression \cite{ Matsui:1986dk,Karsch:1987pv}.  In recent years 
interest has shifted to bound states of bottom and anti-bottom quarks (bottomonium) for the following reasons
\begin{itemize} 
\item
Bottom quarks ($m_b \simeq 4.2$~GeV) are more massive than charm quarks 
($m_c \simeq 1.3$~GeV) and as a result the heavy quark effective theories underpinning 
phenomenological applications are on much surer footing. 
\item
The masses of bottomonium states ($m_\Upsilon \approx$~10 GeV) are much higher than the 
temperatures ($T \simle 1$ GeV) generated in relativistic heavy ion collisions.  As a result,  
bottomonium production will be dominated by initial hard scatterings.
\item
Since bottom quarks and anti-quarks are relatively rare within the plasma, the probability for 
regeneration of bottomonium states through recombination is much smaller than for charm 
quarks.
\item
Due to their higher mass, the effects of initial state nuclear suppression are expected to
be reduced (particularly at central rapidities).
\end{itemize}
As a result one expects the bottomonium system to be a cleaner probe of the quark gluon
plasma than the charmonium system for which the modeling has necessarily become quite
involved.  For this reason I will focus on the bottomonium states in this paper and only consider
the thermal suppression of these states, ignoring initial state effects and any possible thermal
generation or recombination.

In this paper I will review recent theoretical calculations of bottomonium suppression at 
energies probed in relativistic heavy ion collisions at the Large Hadron Collider (LHC).  I will
present a brief overview of the important aspects of the calculation and refer the reader to
Refs.~\cite{Strickland:2011mw,Strickland:2011aa} for details.  I will compare my prediction
for inclusive $\Upsilon(1s)$ and $\Upsilon(2s)$ suppression with recent preliminary data to
be released from the CMS collaboration.  In addition, I present
predictions for $\Upsilon(1s)$ suppression at forward rapidities in 
order facilitate comparison with results forthcoming from the ALICE collaboration.

%---------------------------------------------------------------------------------------------------------------
\section{Theoretical methodology}
%---------------------------------------------------------------------------------------------------------------

In recent years there have been important theoretical developments in heavy quarkonium
theory.  Chief among these are the first-principles calculations of imaginary-valued contributions
to the heavy quark potential.  The first calculation of the leading-order perturbative imaginary 
part of the potential due to gluonic Landau damping was performed by Laine 
et al.~\cite{Laine:2006ns}.  Since then an additional imaginary-valued 
contribution to the potential coming from singlet to octet transitions has also been computed 
using the effective field theory approach~\cite{Brambilla:2008cx}.  These imaginary-valued
contributions to the potential are related to quarkonium decay processes in the plasma.  The 
consequences of such imaginary parts on heavy quarkonium spectral functions 
\cite{Burnier:2007qm,Miao:2010tk}, perturbative thermal widths \cite{Laine:2006ns,%
Brambilla:2010vq}, in a T-matrix approach 
\cite{Grandchamp:2005yw,Rapp:2008tf,Riek:2010py}, and
in stochastic real-time dynamics \cite{Akamatsu:2011se} have recently been studied.  

In addition, there have been significant advances in the dynamical models used to simulate
plasma evolution.  In particular, there has been a concerted effort to understand the effects
of plasma momentum-space anisotropies generated by the rapid longitudinal expansion of
the matter along the beamline direction.  Dynamical models are now able to describe the
anisotropic hydrodynamical evolution in full (3+1)-dimensional simulations 
\cite{Florkowski:2010cf,Martinez:2010sc,Ryblewski:2010bs,Martinez:2010sd,%
Martinez:2012tu,Ryblewski:2012rr}.  
This is important because 
momentum-space anisotropies can have a significant impact on quarkonium suppression
since in regions of high momentum-space anisotropy one expects reduced quarkonium binding
\cite{Dumitru:2007hy,Dumitru:2009ni,Burnier:2009yu,Dumitru:2009fy,%
Margotta:2011ta}.  In Refs.~\cite{Strickland:2011mw,Strickland:2011aa} the dynamical
evolution of the anisotropic plasma was combined with the real and imaginary parts of the binding
energy obtained using modern complex-valued potentials.

%---------------------------------------------------------------------------------------------------------------
%\mysubsec{Potential model and binding energies}
%---------------------------------------------------------------------------------------------------------------

The heavy quark potential has real and imaginary parts, $V = \Re[V] + i \Im[V]$.  Here I will 
focus on a model in which the real part of the potential is obtained from internal energy of the 
system since models based on the free energy seem to be incapable of reproducing either the 
LHC or RHIC  $R_{AA}[\Upsilon]$.  The real part of the potential is given by 
\cite{Strickland:2011aa}
\begin{eqnarray}
\Re[V] =  -\frac{a}{r} \left(1+\mu \, r\right) e^{-\mu \, r }
+ \frac{2\sigma}{\mu}\left[1-e^{-\mu \, r }\right]
- \sigma \,r\, e^{-\mu \, r } -  \frac{0.8\,\sigma}{m_Q^2 r} 
\, , \label{eq:real_pot_model_B}
\end{eqnarray}
where $a=0.385$ and $\sigma = 0.223\;{\rm GeV}^2$ \cite{Petreczky:2010yn} and the
last term is a temperature- and spin-independent finite quark mass correction taken from 
Ref.~\cite{Bali:1997am}.  In this
expression $\mu = {\cal G}(\xi,\theta) m_D$ is an anisotropic Debye mass where ${\cal G}$ 
is a rather complicated function which depends on the degree of plasma momentum-space anisotropy, 
$\xi$, and the angle of the line connecting the quark-antiquark pair with respect to the beamline 
direction, $\theta$, and  $m_D$ is the isotropic Debye mass \cite{Strickland:2011aa}.  In the limit 
$\xi \rightarrow 0$ one has ${\cal G} = 1$.

The imaginary part of the potential $\Im[V]$ is obtained from a leading order perturbative calculation 
which was performed in the small anisotropy limit 
\begin{equation} 
\Im[V] = - \alpha_s C_F T \left\{ \phi(r/m_D) - \xi \left[ \psi_1(r/m_D,
\theta)+\psi_2(r/m_D, \theta)\right] \right\} ,
\label{impot}
\end{equation}
where $\phi$, $\psi_1$, and $\psi_2$ can be expressed in terms of hypergeometric functions 
\cite{Dumitru:2009fy}.

After combining the real and imaginary parts of the potential, the 3d Schr\"odinger equation 
is solved numerically to obtain the real and imaginary parts of the binding energy as a function 
of $\xi$ and $p_{\rm hard}$ \cite{Margotta:2011ta,Strickland:2009ft}.  The imaginary part of 
the binding energy is related to the width of the state
\beq
\Gamma(\tau,{\bf x}_\perp,\varsigma) = 
\left\{
\begin{array}{ll}
2 \Im[E_{\rm bind}(\tau,{\bf x}_\perp,\varsigma)]  & \;\;\;\;\; \Re[E_{\rm bind}(\tau,{\bf x}_\perp,\varsigma)] >0 \\
10\;{\rm GeV}  & \;\;\;\;\; \Re[E_{\rm bind}(\tau,{\bf x}_\perp,\varsigma)] \le 0 \\
\end{array}
\right.
\label{eq:width}
\eeq
where $\varsigma = {\rm arctanh}(z/t)$ is the spatial rapidity.  The value of 10 GeV in the 
second case is chosen to be large in order to quickly suppress states which are fully unbound 
(which is the case when the real part of the binding energy is negative).

%---------------------------------------------------------------------------------------------------------------
%\mysubsec{Dynamical Model}
%---------------------------------------------------------------------------------------------------------------

The dynamical model used gives the spatio-temporal evolution of the typical transverse
momentum of the plasma partons, $p_{\rm hard}(\tau,{\bf x})$, and the plasma 
momentum-space anisotropy, $\xi(\tau,{\bf x})$, both of which are specified in the local
rest frame of the plasma.  The widths obtained from solution of the 3d Schr\"odinger equation
are then integrated and exponentiated to compute the relative number of states remaining 
at a given proper time.  This quantity is then averaged over the transverse plane taking
into account the local conditions in the plasma and weighting by the spatial probability 
distribution for bottomonium production which is given by the number of binary collisions
computed in the Glauber model with a Woods-Saxon distribution for each nucleus.  For the
temporal integration the initial time is set by the formation time of the state in question.

The resulting $R_{AA}$ is a function of the transverse momentum, $p_T$, the rapidity
$\varsigma$, and the nuclear impact parameter $b$.  To compare to experimental results
transverse momentum cuts are applied using a $n(p_T) = n_0 E_T^{-4}$ spectrum.  
In addition, any cuts on the rapidity due to detector acceptance and centrality are applied
as needed.  For details of the dynamical model and $R_{AA}$ computation I refer the 
reader to Ref.~\cite{Strickland:2011aa}.

%---------------------------------------------------------------------------------------------------------------
%\mysubsec{Initial conditions}
%---------------------------------------------------------------------------------------------------------------

For the initial conditions I use a Woods-Saxon distribution for each nucleus and determine 
the transverse dependence of the initial temperature via the third root of the number of 
participants (wounded nucleons). In the spatial rapidity direction I will investigate two possible
temperature profiles:  (a) a broad plateau containing a boost-invariant central region with 
Gaussian limited-fragmentation at large rapidity
\beq
 n(\varsigma) = n_0 \exp(-(|\varsigma|-\varsigma_{\rm flat}/2)^2/2 \sigma_\varsigma^2)
\; \Theta(|\varsigma|-\varsigma_{\rm flat}/2) \, ,
\label{eq:rapplateau}
 \eeq
where $\varsigma_{\rm flat}=10$ is the width of the central rapidity plateau,
$\sigma_\varsigma = 0.5$ is the width of the limited fragmentation tails,
and $n_0$ is the number density at central rapidity \cite{Schenke:2011tv}; and (b) a 
Gaussian motivated by low-energy fits to pion spectra
\begin{equation}
\label{eq:yprofile}
 n(\varsigma) = n_0 \exp(-\varsigma^2/2\sigma_\varsigma^2) 
\quad
{\rm with}
\quad
 \sigma_\varsigma^2=0.64 \cdot 8 \, c_s^2 \ln \left(\sqrt{s_{NN}}/2 m_p\right) /3(1-c_s^4)
 \; ,
 \label{eq:rapgauss}
\end{equation}
where $c_s = 1/\sqrt{3}$ is the sound velocity, $m_p = 0.938$ GeV is the proton mass, 
and $\sqrt{s_{NN}}$ is the nucleon-nucleon center-of-mass energy 
\cite{Strickland:2011aa,Bleicher:2005tb}.
The temperature distribution is given by $T \sim n^{1/3}$.
I note that (\ref{eq:rapplateau})  has the advantage that it has been tuned to successfully
describe the rapidity dependence of the elliptic flow in LHC heavy ion collisions.

%---------------------------------------------------------------------------------------------------------------
\section{Results and Conclusions}
%---------------------------------------------------------------------------------------------------------------

\begin{figure}[t]
\includegraphics[width=0.5\textwidth]{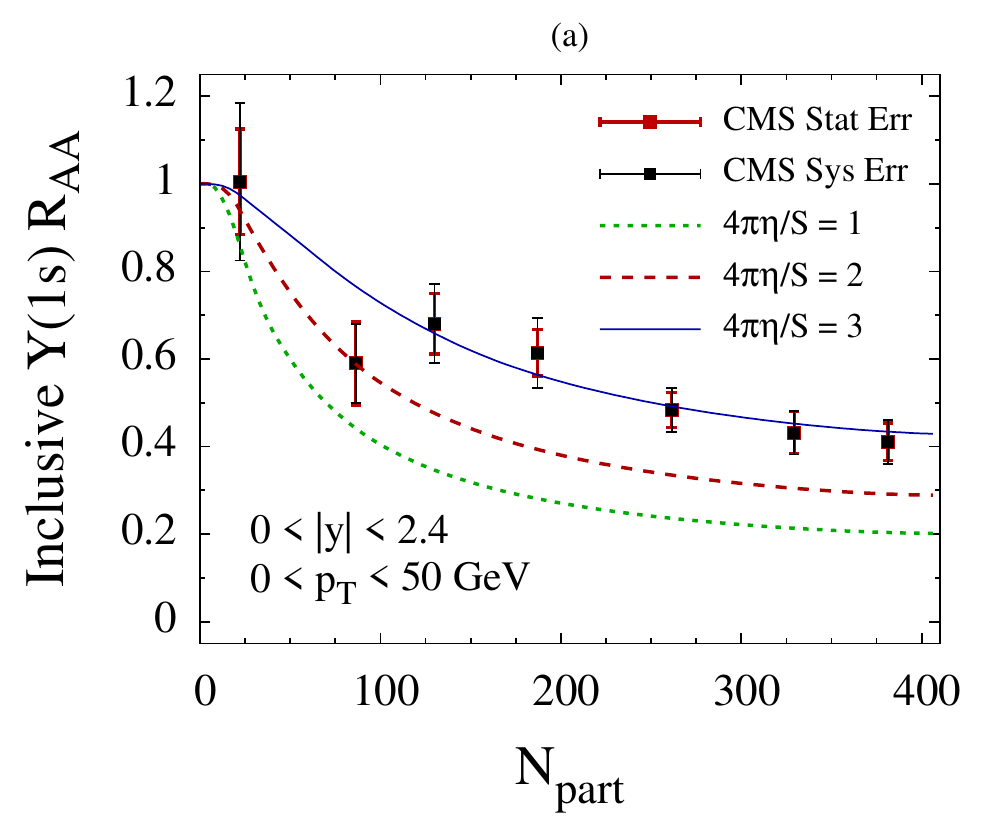}
\hspace{2mm}
\includegraphics[width=0.5\textwidth]{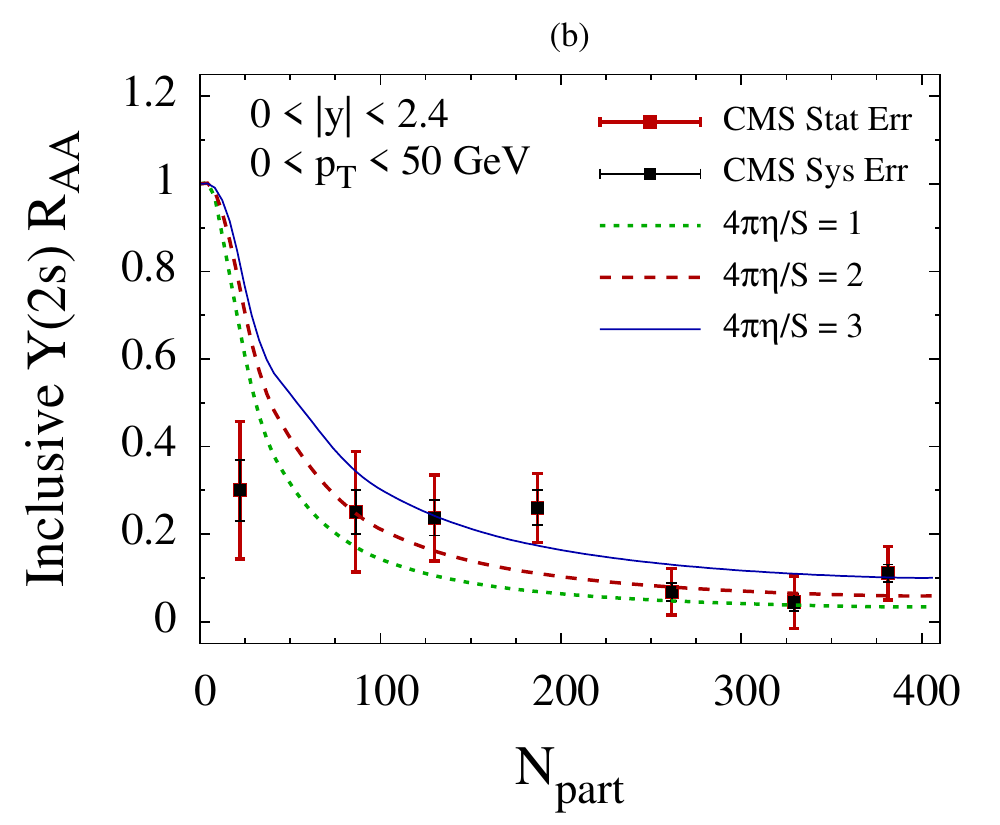}
\caption{Predictions for the central rapidity inclusive (a) $\Upsilon(1s)$ and (b) 
$\Upsilon(2s)$ suppression including feed-down as a function of $N_{\rm part}$ along with 
recent preliminary data from the CMS collaboration.  Three curves show the variation with the 
assumed shear viscosity to entropy ratio.
}
\label{fig:cmscompare}
\end{figure}

When considering the suppression of the $\Upsilon(1s)$ and $\Upsilon(2s)$ states it is 
important to include the effect of feed-down from higher excited states.  In pp collisions 
only approximately 51\% of $\Upsilon(1s)$ states come from direct production and similarly for 
the $\Upsilon(2s)$.  One can compute the inclusive suppression of a state using 
$R_{AA}^{\rm full}[\Upsilon(ns)] = \sum_{i\,\in\,{\rm states}} f_i \,R_{i,AA}$
where $f_i$ are the feed-down fractions and $R_{i,AA}$ is the direct suppression
of each state which decays into the $\Upsilon(ns)$ state being considered.
Here I will use $f_i = \{0.510,0.107,0.008,0.27,0.105\}$ for the $\Upsilon(1s)$, $
\Upsilon(2s)$, $\Upsilon(3s)$,  $\chi_{b1}$, and $\chi_{b2}$ feed-down to $\Upsilon(1s)$, 
respectively \cite{Affolder:1999wm}. For the inclusive $\Upsilon(2s)$ production I use 
$f_i = \{0.500,0.500\}$ for the  $\Upsilon(2s)$ and $\Upsilon(3s)$ states, respectively.
For details of the computation of the direct $R_{AA}$ for each state see 
Ref.~\cite{Strickland:2011aa}.

In Fig.~\ref{fig:cmscompare} I compare to preliminary data on the inclusive $\Upsilon(1s)$ and $\Upsilon(2s)$ suppression available from the CMS collaboration 
\cite{CMS:PhysicsResultsHIN11011}.  For this figure I
used a broad rapidity plateau as the initial density profile as specified in 
Eq.~(\ref{eq:rapplateau}).  The central temperatures 
were taken to be $T_0 = \{520,504,494\}~$MeV at $\tau_0 = 0.3$~fm/c 
for $4 \pi \eta/S = \{1,2,3\}$, respectively, in 
order to fix the final charged multiplicity to $dN_{\rm ch}/dy \simeq 1400$ in each case.  As 
can been seen from this figure, the predictions agree reasonably well with the available
data.  The data seem to prefer the largest value of $\eta/S$ shown; however, there is a
$\pm 14\%$, $\pm 21\%$ 1s, 2s global uncertainty reported by CMS, making
it hard to draw firm conclusions.

\begin{figure}[t]
\includegraphics[width=0.5\textwidth]{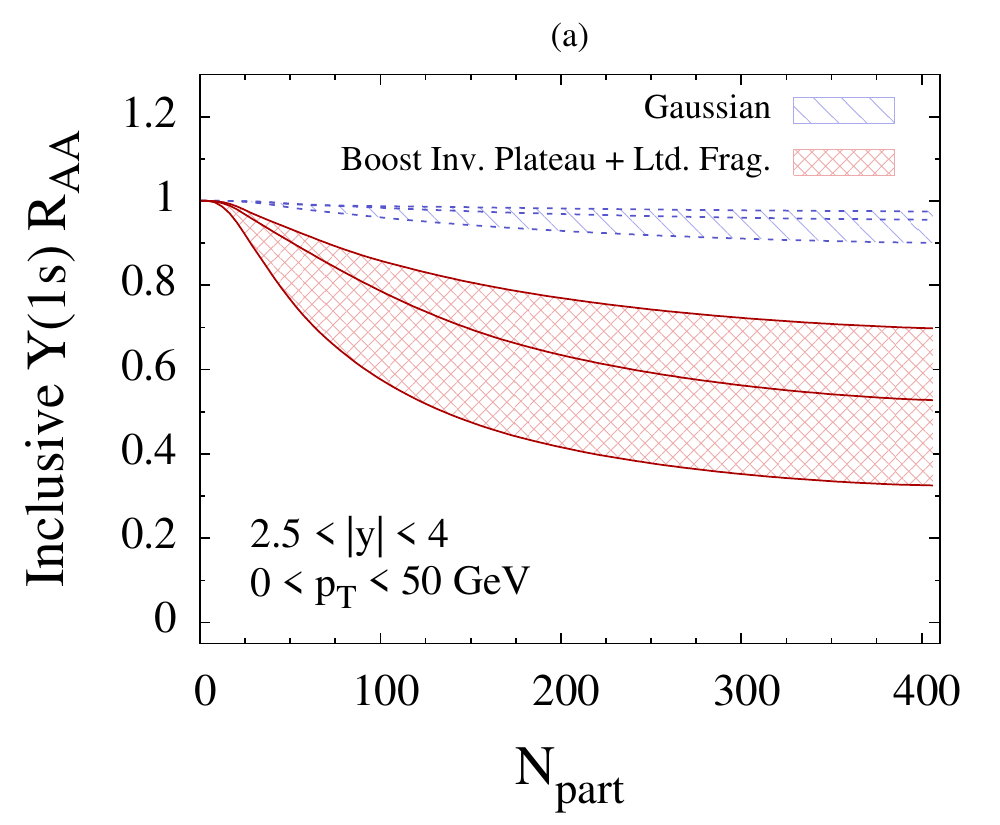}
\hspace{2mm}
\includegraphics[width=0.5\textwidth]{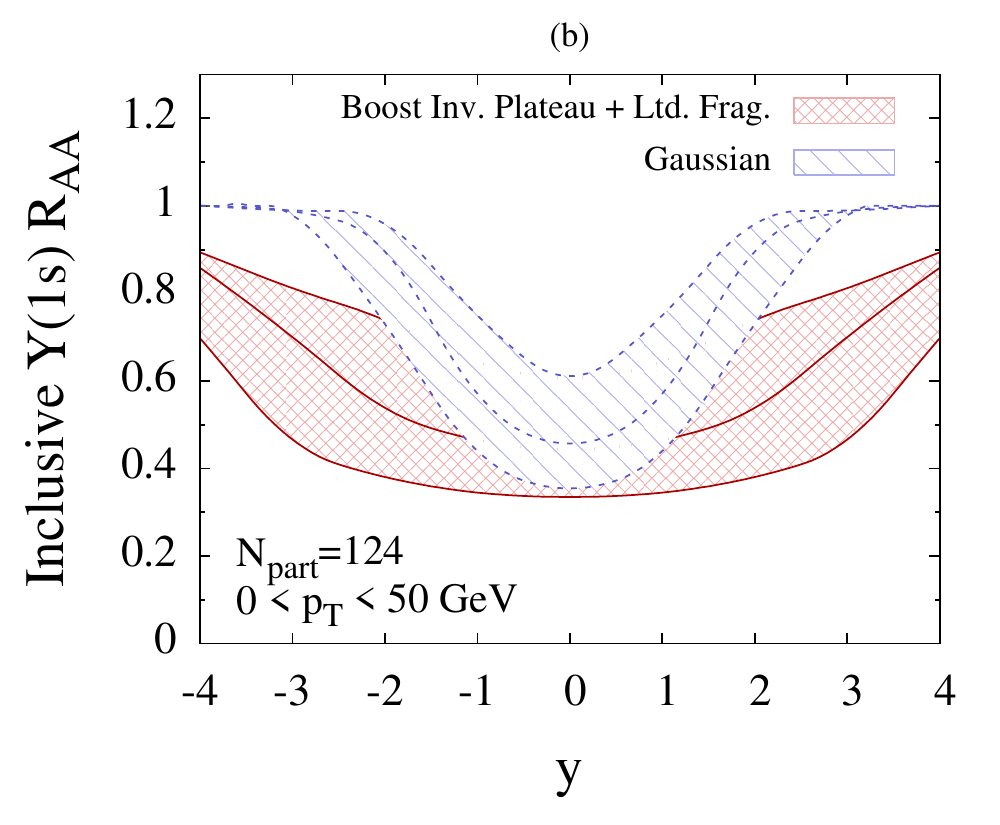}
\caption{Predictions for the inclusive $\Upsilon(1s)$ as (a) a function of $N_{\rm part}$
for $2.5 \leq y \leq 4.0$ and (b) a function of rapidity $y$ for $N_{\rm part} = 124$. 
Bands show variation with $\eta/S$ with the bottom, middle, and top lines in each set 
corresponding to $4\pi\eta/S = \{1,2,3\}$, respectively.
}
\label{fig:aliceprediction}
\end{figure}

Next I consider forward rapidities.  
The ALICE forward muon spectrometer detects charmonium and bottomonium states via
dimuon decay with an acceptance that covers the pseudorapidity
interval $2.5 \leq y \leq 4.0$ with a resolution of approximately 100 MeV.  For more information 
on ALICE's dimuon capabilities see e.g. \cite{Fabjan:2011jb,Das:2011iqa,Das:2011bj,%
Kisslinger:2012tu} and references therein.  One of the key questions that such forward
rapidity measurements will be able to answer is the nature of the temperature distribution
of the plasma at forward rapidities.  

In order to illustrate this in 
Fig.~\ref{fig:aliceprediction} I show predictions for the inclusive suppression of $\Upsilon(1s)$
using either Eq.~(\ref{eq:rapplateau}) or Eq.~(\ref{eq:rapgauss})
as the initial rapidity profile of the density.  In panel (a) my prediction is plotted
as a function of $N_{\rm part}$ with a cut $2.5 \leq y \leq 4$ applied and in panel (b)
my prediction is plotted as a function of $y$ for $N_{\rm part}=124$ which is the average
$N_{\rm part}$ for the 0-90\% centrality bin which will be used by ALICE.
As can be seen from this figure there is a rather large effect from the choice of temperature
profile.  The 
boost-invariant plateau with limited fragmentation (\ref{eq:rapplateau}) 
is the one which is canonically used in heavy-ion models and follows naturally from the 
Bjorken picture, while the Gaussian assumption (\ref{eq:rapgauss}) is more in line with
the Landau picture.  I would expect, based on ability of the Bjorken-like picture to
describe the rapidity dependence of elliptic flow \cite{Schenke:2011tv}, that this is, in 
fact, what one would see in the experiment; however, for completeness sake I show the
Gaussian in order to demonstrate that the future ALICE results in this region of phase space
could easily discern between the two possibilities.  One should note, however, that at forward
rapidities it is expected that CGC and other initial state effects could become important 
(see e.g. Refs.~\cite{Kharzeev:2012py} 
and \cite{Vogt:2010aa}) .
As a result, the purely thermal $R_{AA}$ shown here should be taken as an upper bound of the full
$R_{AA}$ including also initial state effects.

To conclude, in this paper I have attempted to review recent calculations of bottomonium suppression
in LHC heavy ion collisions.  I presented comparisons of prior predictions with 
CMS preliminary data for inclusive $\Upsilon(1s)$ and $\Upsilon(2s)$ suppression.  In addition, I 
have made new predictions for inclusive $\Upsilon(1s)$ suppression at forward rapidities which are
within the ALICE acceptance.  I found that one can easily distinguish between the so-called
Landau and Bjorken pictures of heavy ion collisions with such forward rapidity data.

%%%%%%%%%%%%%%%%%%%%%%%%%%%%%%%%%%%%%%%%%%%%
%% Acknowlegments
%%%%%%%%%%%%%%%%%%%%%%%%%%%%%%%%%%%%%%%%%%%%

\begin{theacknowledgments}
I thank Guillermo Breto Rangel discussions concerning the CMS preliminary results on $\Upsilon(1s)$ 
and $\Upsilon(2s)$ suppression.  I thank Debasish Das for discussions about ALICE's forward 
rapidity capabilities.  This work was supported by NSF grant No. PHY-1068765 and the 
Helmholtz International Center for FAIR LOEWE program.
\end{theacknowledgments}

\bibliographystyle{aipproc}
\bibliography{SE-Strickland}

\begin{thebibliography}{38}
\expandafter\ifx\csname natexlab\endcsname\relax\def\natexlab#1{#1}\fi
\providecommand{\enquote}[1]{``#1''}
\expandafter\ifx\csname url\endcsname\relax
  \def\url#1{\texttt{#1}}\fi
\expandafter\ifx\csname urlprefix\endcsname\relax\def\urlprefix{URL }\fi
\providecommand{\eprint}[2][]{\url{#2}}

\bibitem[Shuryak(1980)]{Shuryak:1980tp}
E.~V. Shuryak, \emph{Phys. Rept.} \textbf{61}, 71--158 (1980).

\bibitem[Matsui and Satz(1986)]{Matsui:1986dk}
T.~Matsui, and H.~Satz, \emph{Phys. Lett.} \textbf{B178}, 416 (1986).

\bibitem[Karsch et~al.(1988)]{Karsch:1987pv}
F.~Karsch, M.~T. Mehr, and H.~Satz, \emph{Z. Phys.} \textbf{C37}, 617 (1988).

\bibitem[Strickland(2011)]{Strickland:2011mw}
M.~Strickland, \emph{Phys.Rev.Lett.} \textbf{107}, 132301 (2011),
  \eprint{1106.2571}.

\bibitem[Strickland and Bazow(2012)]{Strickland:2011aa}
M.~Strickland, and D.~Bazow, \emph{Nucl.Phys.} \textbf{A879}, 25--58 (2012),
  \eprint{1112.2761}.

\bibitem[Laine et~al.(2007)]{Laine:2006ns}
M.~Laine, O.~Philipsen, P.~Romatschke, and M.~Tassler, \emph{JHEP} \textbf{03},
  054 (2007), \eprint{hep-ph/0611300}.

\bibitem[Brambilla et~al.(2008)]{Brambilla:2008cx}
N.~Brambilla, J.~Ghiglieri, A.~Vairo, and P.~Petreczky, \emph{Phys. Rev.}
  \textbf{D78}, 014017 (2008), \eprint{0804.0993}.

\bibitem[Burnier et~al.(2008)]{Burnier:2007qm}
Y.~Burnier, M.~Laine, and M.~Vepsalainen, \emph{JHEP} \textbf{0801}, 043
  (2008), \eprint{0711.1743}.

\bibitem[Miao et~al.(2011)]{Miao:2010tk}
C.~Miao, A.~Mocsy, and P.~Petreczky, \emph{Nucl. Phys.} \textbf{A855}, 125--132
  (2011), \eprint{1012.4433}.

\bibitem[Brambilla et~al.(2010)]{Brambilla:2010vq}
N.~Brambilla, M.~A. Escobedo, J.~Ghiglieri, J.~Soto, and A.~Vairo, \emph{JHEP}
  \textbf{09}, 038 (2010), \eprint{1007.4156}.

\bibitem[Grandchamp et~al.(2006)]{Grandchamp:2005yw}
L.~Grandchamp, S.~Lumpkins, D.~Sun, H.~van Hees, and R.~Rapp, \emph{Phys.Rev.}
  \textbf{C73}, 064906 (2006), \eprint{hep-ph/0507314}.

\bibitem[Rapp et~al.(2010)]{Rapp:2008tf}
R.~Rapp, D.~Blaschke, and P.~Crochet, \emph{Prog.Part.Nucl.Phys.} \textbf{65},
  209--266 (2010), \eprint{0807.2470}.

\bibitem[Riek and Rapp(2011)]{Riek:2010py}
F.~Riek, and R.~Rapp, \emph{New J. Phys.} \textbf{13}, 045007 (2011),
  \eprint{1012.0019}.

\bibitem[Akamatsu and Rothkopf(2012)]{Akamatsu:2011se}
Y.~Akamatsu, and A.~Rothkopf, \emph{Phys.Rev.} \textbf{D85}, 105011 (2012),
  \eprint{1110.1203}.

\bibitem[Florkowski and Ryblewski(2011)]{Florkowski:2010cf}
W.~Florkowski, and R.~Ryblewski, \emph{Phys.Rev.} \textbf{C83}, 034907 (2011),
  \eprint{1007.0130}.

\bibitem[Martinez and Strickland(2010)]{Martinez:2010sc}
M.~Martinez, and M.~Strickland, \emph{Nucl. Phys.} \textbf{A848}, 183--197
  (2010), \eprint{1007.0889}.

\bibitem[Ryblewski and Florkowski(2011)]{Ryblewski:2010bs}
R.~Ryblewski, and W.~Florkowski, \emph{J.Phys.G} \textbf{G38}, 015104 (2011),
  \eprint{1007.4662}.

\bibitem[Martinez and Strickland(2011)]{Martinez:2010sd}
M.~Martinez, and M.~Strickland, \emph{Nucl.Phys.} \textbf{A856}, 68--87 (2011),
  \eprint{1011.3056}.

\bibitem[Martinez et~al.(2012)]{Martinez:2012tu}
M.~Martinez, R.~Ryblewski, and M.~Strickland, \emph{Phys.Rev.} \textbf{C85},
  064913 (2012), \eprint{1204.1473}.

\bibitem[Ryblewski and Florkowski(2012)]{Ryblewski:2012rr}
R.~Ryblewski, and W.~Florkowski, \emph{Phys.Rev.} \textbf{C85}, 064901 (2012),
  \eprint{1204.2624}.

\bibitem[Dumitru et~al.(2008)]{Dumitru:2007hy}
A.~Dumitru, Y.~Guo, and M.~Strickland, \emph{Phys. Lett.} \textbf{B662}, 37--42
  (2008), \eprint{0711.4722}.

\bibitem[Dumitru et~al.(2009{\natexlab{a}})]{Dumitru:2009ni}
A.~Dumitru, Y.~Guo, A.~Mocsy, and M.~Strickland, \emph{Phys.Rev.} \textbf{D79},
  054019 (2009{\natexlab{a}}), \eprint{0901.1998}.

\bibitem[Burnier et~al.(2009)]{Burnier:2009yu}
Y.~Burnier, M.~Laine, and M.~Vepsalainen, \emph{Phys.Lett.} \textbf{B678},
  86--89 (2009), \eprint{0903.3467}.

\bibitem[Dumitru et~al.(2009{\natexlab{b}})]{Dumitru:2009fy}
A.~Dumitru, Y.~Guo, and M.~Strickland, \emph{Phys.Rev.} \textbf{D79}, 114003
  (2009{\natexlab{b}}), \eprint{0903.4703}.

\bibitem[Margotta et~al.(2011)]{Margotta:2011ta}
M.~Margotta, K.~McCarty, C.~McGahan, M.~Strickland, and D.~Yager-Elorriaga,
  \emph{Phys.Rev.} \textbf{D83}, 105019 (2011), \eprint{1101.4651}.

\bibitem[Petreczky(2010)]{Petreczky:2010yn}
P.~Petreczky, \emph{J.Phys.G} \textbf{G37}, 094009 (2010), \eprint{1001.5284}.

\bibitem[Bali et~al.(1997)]{Bali:1997am}
G.~S. Bali, K.~Schilling, and A.~Wachter, \emph{Phys. Rev.} \textbf{D56},
  2566--2589 (1997), \eprint{hep-lat/9703019}.

\bibitem[Strickland and Yager-Elorriaga(2010)]{Strickland:2009ft}
M.~Strickland, and D.~Yager-Elorriaga, \emph{J. Comput. Phys.} \textbf{229},
  6015--6026 (2010), \eprint{0904.0939}.

\bibitem[Schenke et~al.(2011)]{Schenke:2011tv}
B.~Schenke, S.~Jeon, and C.~Gale, \emph{Phys.Lett.} \textbf{B702}, 59--63
  (2011), \eprint{1102.0575}.

\bibitem[Bleicher(2005)]{Bleicher:2005tb}
M.~Bleicher  (2005), \eprint{hep-ph/0509314}.

\bibitem[Affolder et~al.(2000)]{Affolder:1999wm}
A.~A. Affolder, et~al., \emph{Phys.Rev.Lett.} \textbf{84}, 2094--2099 (2000),
  \eprint{hep-ex/9910025}.

\bibitem[{CMS Collaboration}(2012)]{CMS:PhysicsResultsHIN11011}
{CMS Collaboration}, {Observation of $\Upsilon(nS)$ suppression} (2012),
  \urlprefix\url{https://twiki.cern.ch/twiki/bin/view/CMSPublic/PhysicsResultsHIN11011}.

\bibitem[Fabjan and Schukraft(2011)]{Fabjan:2011jb}
C.~Fabjan, and J.~Schukraft  (2011), \eprint{1101.1257}.

\bibitem[Das(2012)]{Das:2011iqa}
D.~Das, \emph{Pramana} \textbf{79}, 863--866 (2012), \eprint{1111.5946}.

\bibitem[Das(2011)]{Das:2011bj}
D.~Das, \emph{Nucl.Phys.} \textbf{A862-863}, 223--230 (2011),
  \eprint{1102.2071}.

\bibitem[Kisslinger and Das(2012)]{Kisslinger:2012tu}
L.~S. Kisslinger, and D.~Das  (2012), \eprint{1207.3296}.

\bibitem[Kharzeev et~al.(2012)]{Kharzeev:2012py}
D.~Kharzeev, E.~Levin, and K.~Tuchin  (2012), \eprint{1205.1554}.

\bibitem[Vogt(2010)]{Vogt:2010aa}
R.~Vogt, \emph{Phys.Rev.} \textbf{C81}, 044903 (2010), \eprint{1003.3497}.

\end{thebibliography}

\end{document}